\documentclass[%
 aip,
 amsmath,amssymb,
 reprint,%
]{revtex4-2}

\usepackage{graphicx}
\usepackage{dcolumn}
\usepackage{bm}

\usepackage[utf8]{inputenc}
\usepackage[T1]{fontenc}
\usepackage{mathptmx}
\usepackage{etoolbox}
\usepackage{nicefrac}

\makeatletter
\def\@email#1#2{%
 \endgroup
 \patchcmd{\titleblock@produce}
  {\frontmatter@RRAPformat}
  {\frontmatter@RRAPformat{\produce@RRAP{*#1\href{mailto:#2}{#2}}}\frontmatter@RRAPformat}
  {}{}
}%
\makeatother
\begin{document}

\preprint{AIP/123-QED}

\title{Wavelength-tunable open double-microcavity to enhance two closely spaced optical transitions}

\author{S. Seyfferle}
 \affiliation{Institut für Halbleiteroptik und Funktionelle Grenzflächen, Center for Integrated Quantum Science and Technology (IQ$^{\text{ST}}$) and SCoPE, University of Stuttgart, Allmandring 3, 70569 Stuttgart, Germany}

\author{T. Herzog}
 \affiliation{Institut für Halbleiteroptik und Funktionelle Grenzflächen, Center for Integrated Quantum Science and Technology (IQ$^{\text{ST}}$) and SCoPE, University of Stuttgart, Allmandring 3, 70569 Stuttgart, Germany}

\author{R. Sittig}
 \affiliation{Institut für Halbleiteroptik und Funktionelle Grenzflächen, Center for Integrated Quantum Science and Technology (IQ$^{\text{ST}}$) and SCoPE, University of Stuttgart, Allmandring 3, 70569 Stuttgart, Germany}

\author{M. Jetter}
 \affiliation{Institut für Halbleiteroptik und Funktionelle Grenzflächen, Center for Integrated Quantum Science and Technology (IQ$^{\text{ST}}$) and SCoPE, University of Stuttgart, Allmandring 3, 70569 Stuttgart, Germany}

\author{S.L. Portalupi}
 \affiliation{Institut für Halbleiteroptik und Funktionelle Grenzflächen, Center for Integrated Quantum Science and Technology (IQ$^{\text{ST}}$) and SCoPE, University of Stuttgart, Allmandring 3, 70569 Stuttgart, Germany}

\author{P. Michler}
 \affiliation{Institut für Halbleiteroptik und Funktionelle Grenzflächen, Center for Integrated Quantum Science and Technology (IQ$^{\text{ST}}$) and SCoPE, University of Stuttgart, Allmandring 3, 70569 Stuttgart, Germany}

\date{\today}

\begin{abstract}
Microcavities have long been recognized as indispensable elements in quantum photonic research due to their usefulness for enhanced light extraction and light-matter interaction.
A conventional high-Q cavity structure typically allows only a single optical transition to be tuned into resonance with a specific mode. The transition to a more advanced double-cavity structure, however, introduces new and interesting possibilities such as enhancing two spectrally close optical transitions at the same time with two distinct cavity modes. 
Here, we investigate a cavity structure composed of a monolithic planar cavity enclosed between two semiconductor distributed Bragg reflectors (DBR) and a top dielectric mirror deposited on a fiber tip. While the bottom cavity is formed by the two DBRs, the mirror on the fiber tip and the top DBR of the semiconductor chip create a second tunable cavity.
These coupled cavities exhibit mode hybridization when tuned into resonance and their splitting can be adjusted to match with the spectral separation of closely spaced optical transitions by a suitable sample design. Furthermore, we report on the simultaneous resonance tuning of the exciton and biexciton transition of a semiconductor quantum dot, each to a separate mode of the open fiber-based double cavity.
Decay time measurements at simultaneous resonance showed a Purcell-factor of $F_P^X$=1.9$\pm$0.4 for the exciton transition.

\end{abstract}

\maketitle

\section{Introduction}
The combination of optical microcavities with semiconductor quantum dots (QDs) looks back on a long-standing success story of achievements in enhanced light extraction efficiency \cite{Unsleber:16}, strong light-matter coupling \cite{stro_cou} and lasing \cite{PM_lased} among others.
The ability of the cavity to influence the spontaneous emission of optical transitions known as Purcell enhancement \cite{purcell} is widely used in cavity quantum electrodynamics to boost the device efficiency and coherence of QD single photon emission and will greatly aid nascent quantum photonic technologies\cite{RevModPhys.87.1379}.
Some of those, notably quantum repeaters\cite{PhysRevLett.98.190503}, make use of entangled photon pairs that can be generated e.g. via the cascaded emission of the biexciton and exciton transitions in a QD \cite{Young_2006, PhysRevLett.96.130501} that possess slightly different emission energies ($\Delta E$=1-3\,meV).
The extraction efficiency and optical quality of these entangled pairs would greatly benefit from cavity-induced Purcell enhancement, however, the simultaneous resonance tuning of both transitions to a cavity mode poses a challenge.

Successful efforts to this end have been made using circular Bragg cavities \cite{best_bullseye, Pan_bullseye_XXX}.
This approach utilizes a spectrally broad cavity mode (Q-factors of 150-300 \cite{bullseye_kola, bullseye_kola_tele}) which enables to encompass exciton and biexciton within the same mode. But due to the spectral separation between exciton and biexciton it is likely that the two lines will be situated at the far side of the Gaussian mode field, thus the light-matter interaction will be reduced as compared to a spectral matching of cavity center wavelength and QD transition.

Another approach \cite{Doppelpillar_senellart} utilized the coupling of a twin micropillar system to the QDs exciton and biexciton and achieved excellent results in Purcell enhanced entangled photon pairs. However, this approach requires very accurate spectral matching of the micropillar modes to the exciton and biexciton transition, respectively, together with spatial matching of the emitter with the cavity field which makes this method difficult to realize.

Here, we utilize an open tunable fiber-based double-cavity \cite{PhysRevB.102.235306} that provides two spectrally separate modes to be tuned into resonance with the QD transitions individually. Additional benefits are the free choice around which emitter to form the resonator as well as considerable fine-tuning possibilities of the cavity length by displacement of the fiber tip with nanometric precision.

\section{Resonance tuning scheme \& cavity design}

\begin{figure}
	\includegraphics[width=1\linewidth]{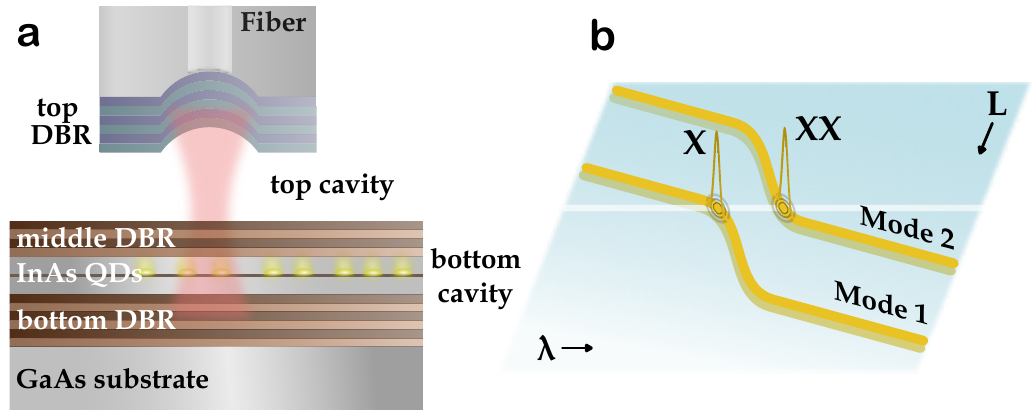}
	\caption{\textbf{Schematic depiction of the double-cavity structure and the simultaneous resonance tuning scheme}. \textbf{a} Cross-section of the fiber double-cavity geometry and the sample structure. \textbf{b} Schematic depiction of the simultaneous resonance tuning scheme of exciton ($X$) and biexciton ($XX$) to longitudinal modes in the anti-crossing region. The spectral distance of two longitudinal modes of the double-cavity structure is reduced to the scale of the binding energy of an exciton/biexciton-pair in the avoided crossing region as the cavity length $L$ is altered.}
	\label{fig:scheme}
\end{figure}

The fiber-based double cavity is constituted of the combination of a planar cavity sample consisting of a $\lambda$-cavity sandwiched between DBRs and a fiber tip placed above the sample surface that serves as the external mirror via its high reflectivity coating. This forms two separate cavities, the monolithic cavity of the sample (\textit{bottom cavity}) and the tunable open cavity (\textit{top cavity}). 
Fig.~\ref{fig:scheme}\textbf{a} illustrates this cavity design.

This geometry allows for the interaction of the modes of each separate cavity that manifests itself in the appearance of the \textit{avoided crossing} phenomenon \cite{PhysRevLett.97.253901, PhysRevB.102.235306} where modes of adjacent mode number approach each other spectrally when tuned into resonance but avoid taking the same energy values due the energy splitting caused by their interaction.
This behaviour is the central element of the scheme to enhance two optical transitions simultaneously since the spectral separation of two longitudinal modes in the avoided crossing region can be engineered by a suitable sample design to be on the order of the spectral separation of an exciton/biexciton-pair for instance \cite{PhysRevB.102.235306}.
The scheme for the dual resonance condition is visualized in Fig.~\ref{fig:scheme}\textbf{b}.
The magnitude of the energy splitting in the avoided crossing region depends on the coupling strength of the involved modes. The larger the coupling strength, the larger the mode separation. In the fiber-based double-cavity this interaction strength is mediated by the respective cavity lengths as well as the reflectivities of the mirrors, but especially by the amount of middle DBR pairs (cf. Fig.~\ref{fig:scheme}\textbf{a}).
An increased number of layer pairs will raise the reflectivity and thereby decrease the mode coupling and their separation as a consequence \cite{PhysRevB.102.235306}.

The basis of the sample grown by metal-organic vapor-phase epitaxy (MOVPE) is a GaAs-substrate onto which the bottom-DBR consisting of 37 pairs of alternating AlAs and GaAs layers is deposited.
This is followed by a $\lambda$-cavity of GaAs containing the active layer of InAs QDs. The structure is completed by 17 AlAs/GaAs DBR-layer pairs, which will act as the middle-DBR in the final double cavity setup. 
The upper cavity is formed by 11 pairs of SiO$_2$/Ta$_2$O$_5$ constituting the high reflectivity coating attached to the fiber end facet. 
The concave imprint on the fiber produced by focused ion beam milling serves as a parabolic resonator mirror and has a curvature radius of 20\,\textmu m \cite{Herzog_2018}. 
The sample is held in a vacuum/helium environment at 4\,K and is attached to piezo actuators to effect its movement relative to the fiber tip in all three directions.
The fiber tip is mounted in a custom made holder for its orthogonal alignment with respect to the sample surface and serves as the excitation as well as the detection path. An above band-gap continuous wave diode laser ($\lambda$=650\,nm) or a pulsed Ti:Sa laser ($\lambda$=800\,nm) for decay time measurements are utilized as excitation sources. 
The QDs embedded in the sample serve as an internal light source of the structure to make the cavity modes visible.

\begin{figure}
	\includegraphics[width=1.00\linewidth]{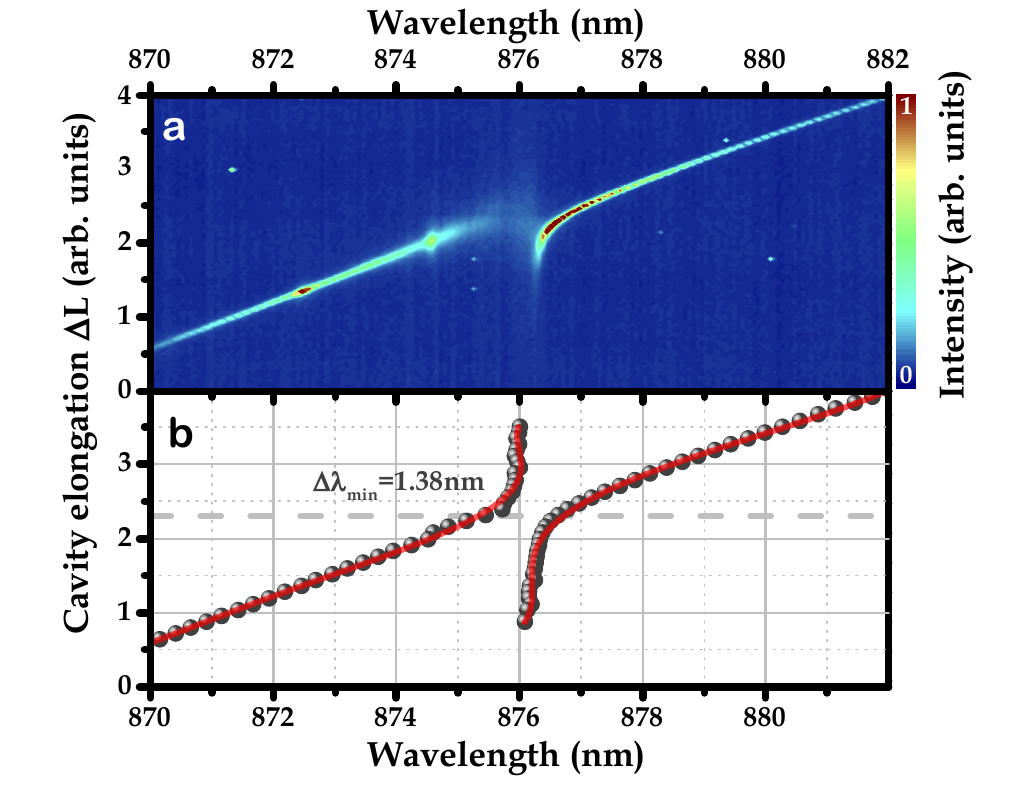}
	\caption{\textbf{Mode separation in the avoided crossing regime}. \textbf{a} Cavity length scan and \textbf{b} extracted mode trajectories for a sample with 17 middle DBR pairs. The data points of the mode trajectories have been evaluated with a polynomial fit in \textbf{b} and their minimum separation is calculated to 1.38\,nm.}
	\label{fig:anti-cross}
\end{figure}

\section{Experimental results}

\subsection{Simultaneous resonance tuning}

As shown in a previous study \cite{PhysRevB.102.235306}, the amount of middle-DBR layer pairs have a distinct influence on the mode separation in the avoided crossing regime, specifically, an increased number of layer pairs reduces the mode splitting. As demonstrated in Ref. 14 the increase from 9 to 12 middle DBR-pairs reduced the minimal mode splitting from 10\,nm to 5.5\,nm.  
Here, a sample of 17 middle DBR pair layers has been fabricated and  used for the formation of the fiber-based double-cavity.
The mode trajectories in dependence of the cavity length $L$ is investigated in Fig.~\ref{fig:anti-cross}\textbf{a}. The cavity length has been continuously altered by stepwise movement of the fiber tip above the sample accompanied by the acquisition of a spectrum at every increment. The resulting scan can be depicted as an intensity map (Fig.~\ref{fig:anti-cross}\textbf{a}) that reveals the mode propagation through the avoided crossing region. 
The modes have been successively fitted with a Gaussian function and their extracted trajectory is evaluated with a polynomial fit in Fig.~\ref{fig:anti-cross}\textbf{b}.
The calculation of the difference of the polynomial fit of the two respective modes yields a minimum separation value of $\Delta \lambda _\text{min}$=1.38\,nm and demonstrates the reduction of the spectral distance between modes of adjacent mode number to align with closely spaced optical transitions in the range of 1-2\,nm.

\begin{figure*}
	\includegraphics[width=1\linewidth]{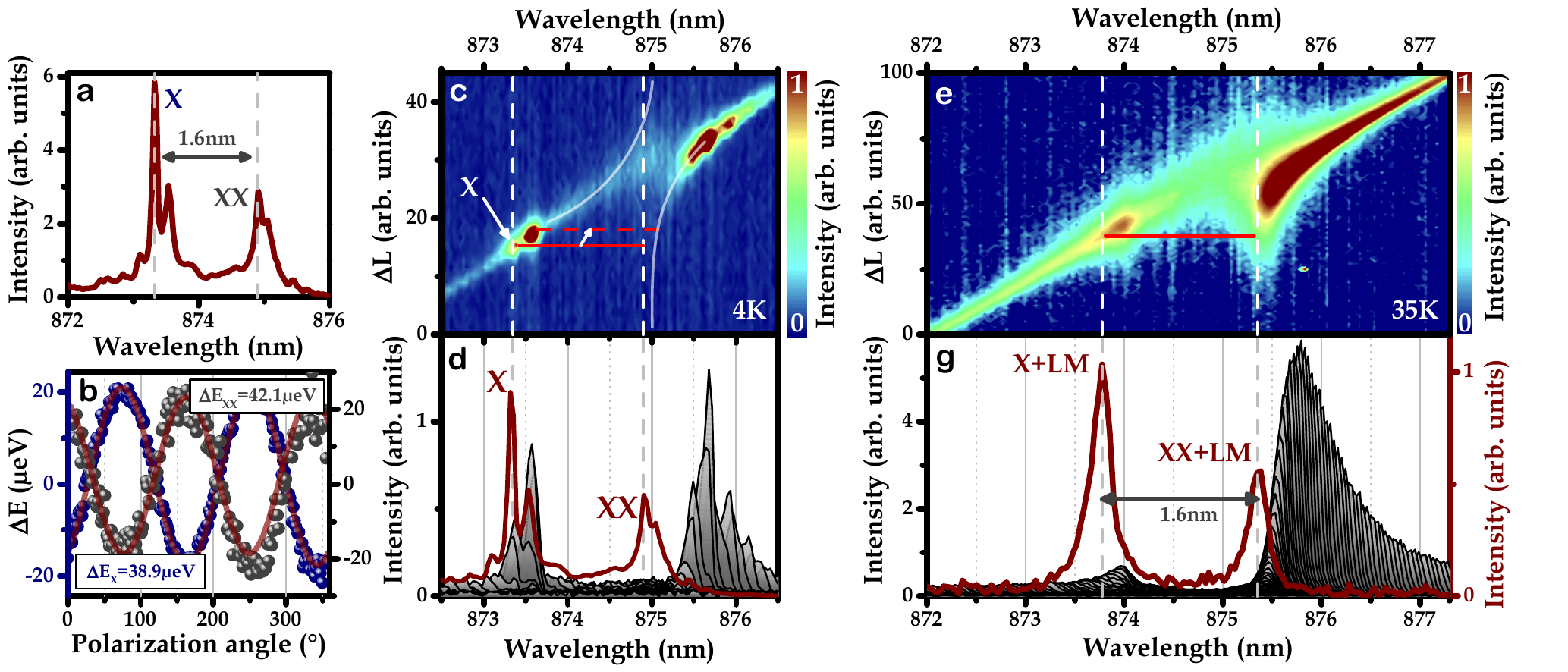}
	\caption{\textbf{Simultaneous resonance tuning of an exciton/biexciton-pair and two modes of the fiber-based double cavity}. \textbf{a} Spectrum of the utilized exciton/biexciton-pair acquired during a pre-selection process step. \textbf{b} Polarization dependent energy measurement of the emission lines designated as $X$ (blue) and $XX$ (black) in \textbf{a}. The anti-phase oscillations and very similar energy splitting of $\Delta E_X$=38.9\,\textmu eV and $\Delta E_{XX}$=42.1\,\textmu eV support the assignment of the transitions as exciton and biexciton of the same QD. \textbf{c} Cavity length scan of the $X/XX$-pair in \textbf{a} in a fiber-based double cavity at a temperature of 4\,K. The appearance of the exciton is designated in the plot, the biexciton comes to lie within the avoided crossing region and is thus not supported by the modes as indicated by the upper red line. Increasing the temperature red-shifts the QD transitions and establishes the simultaneous resonance tuning as visualized by the lower red solid line and the white solid guide to the eye lines roughly representing the mode trajectories. \textbf{d} Comparison of the pre-selection spectrum (red) from \textbf{a} with the scan (grey) from \textbf{c}. \textbf{e} Cavity length scan of the dual resonance configuration at 35\,K. The red bar indicates the cavity length of simultaneous resonance. \textbf{g} Spectrum of the dual resonance in red at the cavity length marked by the red solid line in \textbf{e} superimposed with the stacked spectra of the cavity length scan from sub-figure \textbf{e} indicates the resonance tuning of $X$ and $XX$ each to one of the longitudinal modes ($LM$).}
	\label{fig:reso}
\end{figure*}

In the next step, exciton/biexciton-pairs have been pre-selected via the deposition of marker structures in a deterministic low-temperature optical lithography process \cite{insitu_pill_senellart, Marc_2nm}. This allows for the later retrieval of the selected emitter in the fiber cavity set-up by locating the markers via an in-plane reflectivity scan \cite{Herzog_2018}.
Selection requirements are the appearance of the emission lines in question in the wavelength range of the anti-crossing region as well as their splitting of $\sim$1.4\,nm.
A suitable choice is depicted in Fig.~\ref{fig:reso}\textbf{a}. The spectral distance of the transitions marked as $X$ and $XX$ in the graph amounts to 1.6\,nm and Fig.~\ref{fig:reso}\textbf{b} indicates the nature of these emission lines being exciton and biexciton by their anti-phase peak oscillations under polarization dependent energy detection \cite{pol_serie}.
The fiber-based double cavity is then formed around the designated QD and a cavity length scan is conducted at 4\,K to investigate the spectral relations of the exciton/biexciton transitions and the anti-crossing. 
The result is displayed in Fig.~\ref{fig:reso}\textbf{c}.
One of the intensity spots (marked in the plot as $X$) can be identified as the exciton of the pre-selected QD with the help of Fig.~\ref{fig:reso}\textbf{d}. Here, the scan from sub-figure \textbf{c} is reduced to one dimension by stacking every single spectrum on top of each other and superimposed with the pre-selection spectrum from Fig.~\ref{fig:reso}\textbf{a}. The biexciton happens to be located within the intensity depleted region of the anti-crossing and is therefore not supported by mode enhancement as indicated by the upper red line in Fig.~\ref{fig:reso}\textbf{c}. 
Increasing the temperature red-shifts the transitions and moves the biexciton into the cavity mode at the higher wavelength side of the avoided crossing resulting in a situation represented by the lower red line and the white lines as guides to the eye for the mode trajectories in Fig.~\ref{fig:reso}\textbf{c}. 
The simultaneous resonance condition can be established at 35\,K as depicted in Fig.~\ref{fig:reso}\textbf{e}. The mode emission intensity has increased due to the higher rate of phonon-assisted cavity feeding at elevated temperatures.
Due to this broad enhancement singular spots that would indicate the QD transitions in the cavity cannot be distinguished in Fig.~\ref{fig:reso}\textbf{e}.
The solid red line drawn on top of the data in Fig.~\ref{fig:reso}\textbf{e} marks the position of the exciton at the left end of the line and the position of the biexciton on the other end. 
Fig.~\ref{fig:reso}\textbf{g} gives a comparison of the stacked plots of the scan of Fig.~\ref{fig:reso}\textbf{e} and the spectrum taken at the simultaneous resonance at the cavity length marked by the red line in sub-figure \textbf{e}. 
 The spectrum shows two separate emission lines representing the longitudinal mode (LM) including the exciton emission at lower wavelengths and the biexciton emission in the corresponding other mode. The peak separation is 1.6\,nm and corresponds exactly to the spectral distance of exciton and biexciton as seen in Fig.~\ref{fig:reso}\textbf{a}. In order to demonstrate the dominant contribution of exciton and biexciton emission to the mode emission an analysis of the decay dynamics of the modes ($X+LM$, $XX+LM$) is discussed in the following.

\subsection{Decay dynamics}
\begin{figure}
	\includegraphics[width=1\linewidth]{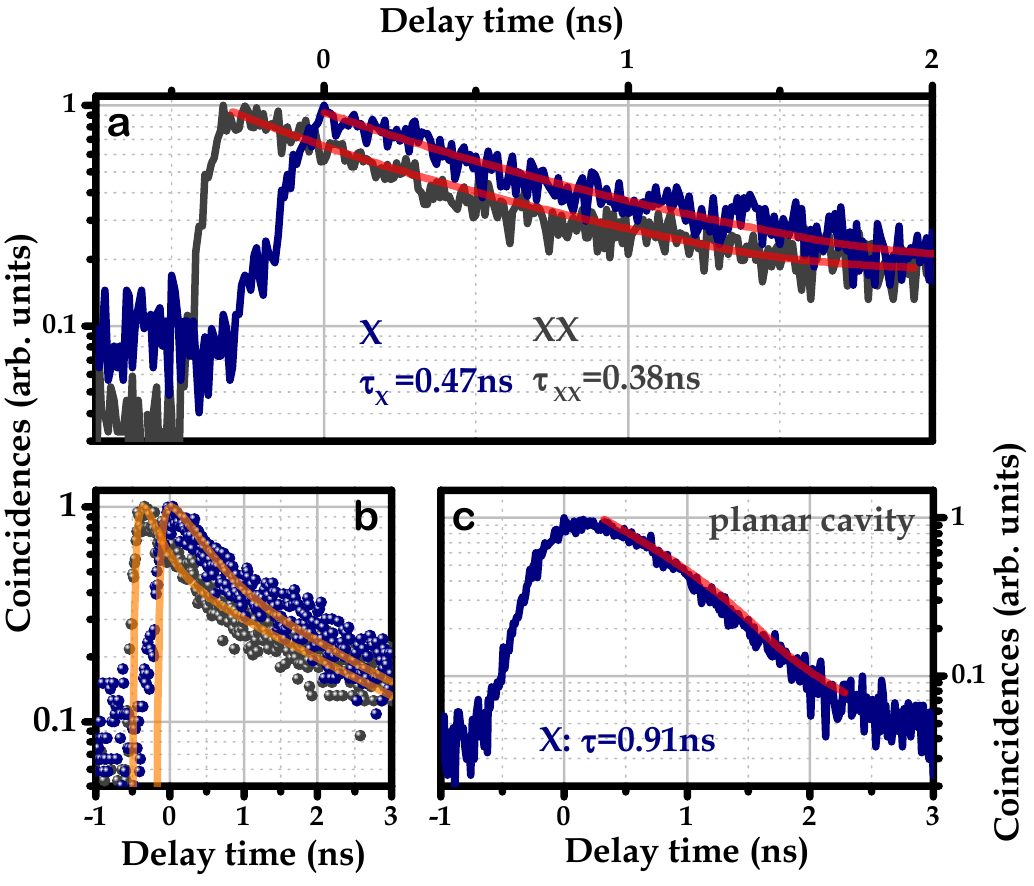}
	\caption{\textbf{Decay dynamics at the simultaneous resonance condition.} \textbf{a} TCSPC measurement of the exciton (blue) and biexciton (black) each simultaneously tuned to resonance with a cavity mode on the basis of the spectrum of Fig.~\ref{fig:reso}\textbf{g}. The resulting decay constants of an exponential fit (red curves) amount to $\tau _X$=0.47$\pm$0.07\,ns and $\tau _{XX}$=0.38$\pm$0.09\,ns. \textbf{b} The measured decay traces are compared to numerical solutions of a rate equation model (see supplementary) based on the fit results from sub-figure \textbf{a}. The simulations (orange curves) are well in accordance with the data, thus supporting the assumed exciton/biexciton decay behaviour. \textbf{c} TCSPC measurement of the exciton transition in the planar cavity environment to estimate the Purcell-factor $F_P$. The decay time is extracted to $\tau ^X_\text{plan}$=0.91$\pm$0.07\,ns.}
	\label{fig:tcpc}
\end{figure}

Based on the dual resonance configuration the decay dynamics have been measured via time-correlated single photon counting (TCSPC). The resulting histogram is plotted in Fig.~\ref{fig:tcpc}\textbf{a}. It can be noted that the increase of the exciton coincidences occurs with a time delay in relation to the rise of the biexciton coincidences and is less steep in comparison. After the biexciton decay sets in the exciton follows, which is an indicator of the cascaded emission typical for these two transitions.
An exponential fit to the data after deconvolution with the instrument response function yielded decay time constants of $\tau _X$=0.47$\pm$0.07\,ns and $\tau _{XX}$=0.38$\pm$0.09\,ns for the exciton and biexciton transitions, respectively.
The ratio of the decay times is unlike the expected 2:1 relations that exciton and biexciton usually exhibit. It appears that the decay times are modified in a way that the exciton recombination is enhanced at a higher rate than the biexciton decay. This points towards the fact that each transition is subject to a different cavity environment that accelerates the decay at a different rate. Since both measurements have been made at the same cavity elongation the mode volume is the same for both. So, the reason has to be found in the non-linear dependence of the Q-factor on the cavity elongation as simulations in a preceeding study \cite{PhysRevB.102.235306} showed.
Thus, it can be argued that the biexciton is resonant to a segment of the mode where the Q-factor is smaller as compared to the Q-factor at the spectral position of the exciton and consequently, the exciton transition is enhanced at a higher rate.

A treatment of the data with a rate equation model (detailed in the supplementary material) suggests itself in order to confirm the nature of the exciton-biexciton dynamic in the decay behaviour. A numerical solution of the model convolved with the instrument response function using the parameters obtained from the fit in Fig.~\ref{fig:tcpc}\textbf{a} ($\tau _X$=0.47\,ns, $\tau _{XX}$=0.38\,ns) results in the decay curves plotted as orange solid lines in Fig.~\ref{fig:tcpc}\textbf{b}.
A secondary decay time constant of $\tau _2$=8\,ns inserted into the model has been found necessary to best reproduce the decay traces. This hints at the influence of the elevated temperature of 35\,K which likely causes refilling by a nearby trap state.
The comparison of the calculated decay curves (solid orange lines) with the measured decay dynamics displayed as data points in blue (exciton) and black (biexciton) in Fig.~\ref{fig:tcpc}\textbf{b} shows a very good agreement between model and measurement. This underlines the assumption of a dominating exciton/biexciton decay behaviour in the measurement.

An estimation of the Purcell-factor achieved in the simultaneous resonance configuration requires the knowledge of the bare decay time.
To this end, the exciton transition has been measured without a fiber installed to obtain its decay time in the planar cavity. The sample temperature was fixed to 35\,K to grant a direct comparison to the conditions of Fig.~\ref{fig:reso}\textbf{g}.
The resulting histogram is seen in Fig.~\ref{fig:tcpc}\textbf{c}.
The decay constant of the exciton transition can be determined with an exponential fit after deconvolution and amounts to $\tau _{\text{plan}}^X$=0.91$\pm$0.07\,ns.
Note, that the biexciton decay could not be measured at 35\,K due to an encountered strong intensity depletion of the emission line at high temperatures which is why a reasonable estimation of the Purcell-factor of the biexciton transition is not provided. 
The ratio of the decay times in the planar and double cavity environment of the exciton transition allows to estimate the Purcell-factor which amounts to $F_P^X$=1.9$\pm$0.4.

\section{Conclusions}

In summary, a study of the properties of a tunable fiber-based double-cavity in view of supporting two closely spaced optical transitions via two separate modes has been presented. The sample design can be engineered with respect to mode interaction to enable a mode separation in the anti-crossing regime on the order of $\sim$1-2\,nm.
A realization of the simultaneous resonance tuning has been shown with the exciton and biexciton transitions of a semiconductor QD individually brought to resonance with either one of the modes.
A first estimation of the Purcell-factor under these conditions amounts to $F_P^{X}$=1.9$\pm$0.4 for the exciton transition.
Future progress should involve the implementation of strain tuning to establish the simultaneous resonance condition without having to rely on temperature tuning. Additionally, strain tuning can modify the biexciton binding energy which would provide more flexibility in terms of spectral matching of the QD transitions and the cavity modes .

\begin{acknowledgments}

We gratefully acknowledge the support from DFG MI 500/31-1. Parts of this work were financially supported by the program Competence Network Quantum Technology in BW (QTBW.net) from the Baden-Württemberg Stiftung. Furthermore, we thank L. Engel and S. Kolatschek for their support in the clean-room and sample processing steps and F. Hornung for the help in data evaluation.
\end{acknowledgments}




\section*{References}
\bibliography{aipsamp}

\end{document}